# Weak Localization and Anomalous Hall Effect in Chemically Disordered $L1_0$-$Mn_{1.5}Ga$

L. J. Zhu, D. Pan, and J. H. Zhao*

*State Key Laboratory of Superlattices and Microstructures, Institute of Semiconductors, Chinese Academy of Sciences, P. O. Box 912, Beijing 100083, China*

**Abstract:** The transport behaviors of weak localization and anomalous Hall effect (AHE) in perpendicularly magnetized $L1_0$-$Mn_{1.5}Ga$ single-crystalline films are investigated as a function of degree of long-range chemical ordering and temperature. We observed significant weak localization and metal-insulator transition in highly disordered films. Our results provide new evidence that weak localization, phonons and magnons have negligibly smaller effect on extrinsic AHE resistivity than static defects for all the chemical disordered films, the overlook of which in conventional scaling laws may result in significant discrepancies and exponent $n$ beyond 2 when fitting the data. Chemical disorder is found to exhibit strong tailoring effect on both intrinsic and extrinsic $\sigma_{AH}$. Our results clarify the role of weak localization, phonons and magnons in AHE and offer a technological engineering approach of the AHE by altering long-range chemical ordering parameter of magnetic materials.

[a] Author to whom correspondence should be addressed;

Electronic mail: jhzhao@red.semi.ac.cn

The anomalous Hall effect (AHE) has attracted much attention due to both the fascinating underlying physics and the great application potential in sensors, memories and logics.[1-8] It is widely accepted that there are three mechanisms responsible for the AHE. The intrinsic AHE arises from the transverse velocity of the Bloch electrons induced by SOI and interband mixing,[9] as experimentally evidenced by quantum anomalous Hall effect.[10] The extrinsic mechanisms of skew scattering and side jump result from the impurity scattering of conduction electrons due to SOI.[11,12] Skew scattering yields a scaling relation between longitudinal resistivity $\rho_{xx}$ (conductivity $\sigma_{xx}$=1/$\rho_{xx}$) and anomalous Hall resistivity $\rho_{AH}$ (conductivity $\sigma_{AH}\approx\rho_{AH}/\rho_{xx}^2$) as $\rho_{AH}\sim\rho_{xx}$, while other two mechanisms give $\rho_{AH}\sim\rho_{xx}^2$. Accordingly, scaling laws $\rho_{AH}\sim\rho_{xx}^n$ and $\rho_{AH}=a\rho_{xx}+b\rho_{xx}^2$ are conventionally used to describe the experimental data.

In spite of the extensive studies, a unified scaling of AHE has remained a shortage. In stark contrast to present AHE theories, the exponent $n$ was experimentally observed to be below 1 or beyond 2 in a variety of ferromagnets, e.g. 0.4 in hopping transport-accompanied $Fe_3O_4$ films,[13] 2.7 (2.85) in (111)-orientated $Mn_{2.03}Ga$ ($Mn_{1.38}Ga$) films,[14] 2.6 in Fe/Cr multilayers[15] and 3.7 in Co-Ag granular films.[16] Electron localization was recently found to yield a scaling with $n$=0 in chemically disordered FePt ultrathin films.[17] Noticeably, recent studies in Fe films revealed a negligible contribution of phonon skew scattering and a scaling $\rho_{AH}=a_0\rho_{xx0}+b\rho_{xx}^2$, where $a_0\rho_{xx0}=\alpha_0\rho_{xx0}+\beta\rho_{xx0}^2$ is the extrinsic contributions of skew scattering and side jump from static defects.[3] These intriguing observations make the scaling of AHE open questions. On the other hand, chemical disorder and composition were claimed to have influence on intrinsic AHE by affecting the topology of Fermi surface, as suggested in alloys of FePt, $FePt_xPd_{1-x}$, $Co_2Fe_xMn_{1-x}Si$, $CoFeSi_{0.6}Al_{0.4}$ and $CoFeGa_{0.5}Ge_{0.5}$ films,[5,18-20] although it is always believed that intrinsic AHE originates from the band effects of perfect crystal and is independent of the disorders.[1,2,9,21] The exploration and clarification of these puzzling and intriguing issues is of fundamental importance for better understanding the AHE and underlying physics, and also for technological applications in sensors, memories and logics.[18]

Alloys of $L1_0$-$Mn_{1.5}Ga$ have great application potential in ultrahigh-density perpendicular magnetic recording, high-performance spintronic devices and economical permanent magnets due to their giant perpendicular magnetic anisotropy, large magnetic energy product, ultrahigh coercivity and ultralow magnetic damping constant.[22] $L1_0$-$Mn_{1.5}Ga$ films are also ideal to systematically explore the effects of electron localization and chemical disorder on the AHE, as their perfectly square hysteresis allow for highly accurate determination of $\rho_{AH}$, and the chemical disorder strength could be controllably altered for tailoring the related transport behaviors.[23]

In this paper we study the transport behaviors in perpendicularly magnetized $L1_0$-$Mn_{1.5}Ga$ films as a function of degree of long-range chemical ordering ($S$) and temperature ($T$). We observed significant quantum transport of weak localization in highly disordered films. The scaling of AHE in all the differently disordered films could be excellently described by $\rho_{AH}=a_0\rho_{xx0}+b\rho_{xx}^2$ instead of the conventional laws, suggesting the negligible contribution to extrinsic $\rho_{AH}$ and great suppression of extrinsic $\sigma_{AH}$ at finite $T$ by weak localization, phonons and magnons. Chemical disorder exhibits strong tailoring effect on both intrinsic and extrinsic $\sigma_{AH}$. Our results clarify the role of weak localization, phonons and magnons in AHE and offer a technological engineering approach of the AHE by altering long-range chemical ordering parameter of magnetic materials.

A series of 50 nm $L1_0$-$Mn_{1.5}Ga$ single-crystalline films were epitaxially grown on semi-insulating GaAs (001) substrates by molecular-beam epitaxy respectively at 100,150, 200, 250 and 300 °C to tune the long-range chemical ordering parameter $S$, and then capped with a 6 nm-thick GaAs layer to prevent oxidation.[22] The details of sample fabrication, structural characterizations and magnetic properties could be found elsewhere.[23] The good homogeneity and sharp interface of these films are confirmed by the streaky reflection high-energy electron diffraction (RHEED) patterns (Fig. 1(a)), the strongly oscillatory x-ray reflection (XRR) curves (Fig. 1(b)) and transmission electron microscopy images.[23] The films were patterned into 60×300 $\mu m^2$ Hall bars using ultraviolet photolithography and ion-beam etching for transport measurements. Hall resistivity $\rho_{xy}$ was measured by physical property measurement system (PPMS) together with longitudinal resistivity $\rho_{xx}$. The magnetoresistance in these films are smaller than 0.5% even at 7 T.

Figure 2(a) shows Hall resistivity $\rho_{xy}$ at 300 K as a function of perpendicularly applied magnetic field $H$ for all the MnGa films with different $S$. The variation of coercivity in Hall hysteresis loops reveals the different disorder strength in these films as discussed in Ref. 23. The anomalous Hall resistivity $\rho_{AH}$ was extrapolated to $H$=0 T from $\rho_{xy}$-$H$ curves up to ±7 T after subtracting the ordinary Hall contribution $R_0H$ following the empirical relationship of $\rho_{xy} = R_0H+R_sM$ with $\rho_{AH}=R_sM$. Figure 2(b) demonstrates typically the $\rho_{AH}$ hysteresis of MnGa films measured at various temperature from 2 to 350 K. The perfectly square $\rho_{AH}$ hysteresis loops reveal the strong perpendicular anisotropy, consistent well with the magnetic hysteresis curves.[23] As summarized in Fig. 2(c), $\rho_{AH}$ increases monotonically with $T$ for each film and shows a large value of 2.5 μΩ.cm at 350 K for the film with $S$ = 0.37, which is still smaller than the reported 11.5 μΩ.cm at room temperature in 30 nm $D0_{22}$-$Mn_2Ga$ films grown on MgO.[24] It should be mentioned that the magnetization and coercivity of each film changes little in the studied temperature range of 2-350 K as revealed by detailed superconducting quantum interference device (SQUID) and PPMS measurements, because of the high Curie temperature,[23] which suggests that influence of $T$-dependence of saturation magnetization on $\rho_{AH}$ could be neglected.[5-7] Figure 2(d) shows the $T$-dependence of $\rho_{xx}$ for MnGa samples with different $S$, revealing that all the investigated samples are metallic above 20 K. As shown in Fig. 2(e), with $T$ further decreases toward 2 K, $\Delta\rho_{xx}$ (=$\rho_{xx}$-$\rho_{xx0}$, $\rho_{xx0}$ is the minimum of $\rho_{xx}$ as shown in the inset) decreases monotonically for each film with $S \geq 0.26$, whereas metal-insulator transition appears at around 20 (12) K in strongly chemically disordered $Mn_{1.5}Ga$ with $S$=0.20 (0.25). $\rho_{xx}$ changes linearly with ln $T$ in the insulator-like regime. A plausible interpretation of this is two-dimensional (2D) weak localization where quantum interference of electron waves diffusing around disorders enhances backscattering and thus leads to a reduction of the conductivity.[25] In effective 2D case, the weak localization-caused quantum correction to sheet conductivity $\Delta G_\square$ could be expressed as $\Delta G_\square = pD_{00}\ln(T/T_0)$, where $D_{00}=e^2/\pi h$ is the conductivity quantum, $T_0$ is related to the transport mean free path, and $p$ is the temperature exponent of the inelastic scattering length $l_{in} \sim T^{p/2}$. Theoretically, the value of $p$ is governed by the inelastic relaxation mechanisms, i.e. $p$=1/2 for electron-magnon scattering;[26] $p$=1 for electron-electron collision; $p$=3 for electron-phonon scattering.[25] Figure 2(f) plots the corrections to sheet conductivity $\Delta G_\square$ as a function of $T$ for $Mn_{1.5}Ga$ with $S$=0.20 (0.25). The nice logarithmic behavior for the data below 15 (8) K verifies the validity of the 2D assumption. The best fitting of the $\Delta G_\square \sim \ln T$ to the logarithmic part yields a slope $p$ of 2.2 (1.3), suggesting a weak localization correction to the conductivity from dominate inelastic relaxation mechanism of electron-electron collision mediated electron-phonon scattering (electron-phonon scattering mediated electron-electron collision) for $Mn_{1.5}Ga$ with $S$ = 0.2 (0.25). This weak localization correction could be also verified by the negative magnetoresistance (MR, Fig. 2 (g)) as a uniform magnetic field destroys the phase coherence of the "closed path" electron path and hence suppressing weak localization effects.[25-26] The negative MR is also the evidence for the fact that the low temperature resistivity is not dominated by the effective electron-electron interaction, which would lead to a logarithmic increase of the resistivity and positive MR.[25]

To distinguish the various contributions to the experimental anomalous Hall resistivity, we first checked the scaling of AHE in $Mn_{1.5}Ga$ following the conventional laws $\rho_{AH} \sim \rho_{xx}^{\ n}$ and $\rho_{AH}=a\rho_{xx}+b\rho_{xx}^2$. Figure 3(a) shows a double-logarithmic plot of $\rho_{AH}$ against $\rho_{xx}$ for MnGa with different $S$, which yields the exponent $n$ by linear fit of $\log_{10}\rho_{AH}$ against $\log_{10}\rho_{xx}$. Noticeably, MnGa with $S$=0.25 shows an exponent $n$ as large as 2.13, similar to the reported value of 2.85 and 2.7 for (111)-orientated $Mn_{1.38}Ga$ and $Mn_{2.03}Ga$ grown on GaN.[14] It is worth mentioning that we also found $n$ larger than 2 in $Mn_{1.5}Ga$ samples with other composition. Although the situation of $n$>2 has been also found in Fe/Cr multilayers with interfacial scattering and 3.7 in Co-Ag granular films with boundary scattering,[15,16] it is rather confusing that so large $n$ is found in our homogenous single-crystalline $Mn_{1.5}Ga$ films as thick as 50 nm. More importantly, one cannot neglect the significant deviations of experimental data from fitting by the single power scaling for the $Mn_{1.5}Ga$ film with $S$>0.26, indicating the invalidity of the data evaluation following the single power law. Figure 3(b) presents the measured $\rho_{AH}/\rho_{xx}$ ratio plotted versus $\rho_{xx}$ and its linear $a+b\rho_{xx}$ fits for the differently disordered samples. Noticeably, all the samples show evident deviation from the linear trend, especially for data at low temperature. When further comparing the $\rho_{AH}/\rho_{xx}$ and $\rho_{xx}$ data at 2K for all these samples, one can also find remarkable nonlinearity, which reveals the different scaling for differently disordered films, and disobeys the widely accepted picture that the test of the scaling relation by fitting $\rho_{AH}/\rho_{xx}$ and $\rho_{xx}$ data in a set of samples with different disorders or defects concentration is always the shortcut to identify the AHE mechanisms.[1,16]

Generally speaking, in bad metal ($\rho_{xx}$>$10^2$ μΩ.cm),[1] static defect scattering, weak localization, electron-electron interaction or hopping conduction could be important, and even give rise to a resistivity minimum (so-called metal-insulator transition). With $T$ further increases, scattering of conduction electrons will be dominated by electron-phonon and electron-magnon scattering. Accordingly, for chemically disordered $Mn_{1.5}Ga$ films, the resistivity could be written as $\rho_{xx}=\rho_{xx0}+\rho_L+\rho_p+\rho_m$, where $\rho_{xx0}$ is residual resistivity arising from defect scattering, $\rho_L$, $\rho_p$ and $\rho_m$ stand for the contributions from weak localization, phonon and magnon scattering, respectively. Skew scattering is conventionally included into AHE in a unified way of $\rho_{sk}=\alpha\rho_{xx}$. However, it has been both theoretically and experimentally argued that defect-induced skew scattering contribution to AHE could be dominating compared to that induced by phonons and magnons.[3,27] Recent experiments also suggested that magnons always give much smaller contribution to AHE in type of skew scattering than phonons.[28] Therefore, it is reasonable to consider the qualitative difference in various contributions to skew scattering and describe AHE resisitivity as $\rho_{AH}=\alpha_0\rho_{xx0}+\alpha_L\rho_L+\alpha_p\rho_p+\alpha_m\rho_m+\beta\rho_{xx0}^2+b\rho_{xx}^2=a_0\rho_{xx0}+\alpha_L\rho_L+\alpha_p\rho_p+\alpha_m\rho_m+b\rho_{xx}^2$, in which the linear terms of $\alpha_0\rho_{xx0}$, $\alpha_L\rho_L$, $\alpha_p\rho_p$ and $\alpha_m\rho_m$ accout for the skew scattering contributions from defects, weak localization, phonons and magnons, respectively, whereas the quadratic term $\beta\rho_{xx0}^2$ and $b\rho_{xx}^2$ for scattering potential-independent side jump and intrinsic contributions.[3] Here we used different footing for different contributions to skew scattering to distinguish their difference. $\beta$ and $b$ are considered as a constant for each film as side jump and intrinsic contribution are always independent of temperature. Noticeably, in case of negligible skew scattering contributions from weak localization, phonons and magnons ($\alpha_L\rho_L=\alpha_p\rho_p=\alpha_m\rho_m=0$), the scaling law will be simplified to $\rho_{AH}=a_0\rho_{xx0}+b\rho_{xx}^2$ with $a_0=\alpha_0+\beta\rho_{xx0}$, which successfully described the AHE scaling in Fe thin films.[3] The equation gives $\rho_{AH}=b\rho_{xx}^2$ for $\rho_{xx0}<<\rho_{xx}$, $\rho_{AH}=a\rho_{xx0}$ for $\rho_{xx}\approx\rho_{xx0}$ (ultralow $T$) and $\rho_{xx0}^2<<\rho_{xx}^2$ (ultrapure samples). As plotted in Fig. 3(c), $\rho_{AH}$ show an excellent linear correlation of $\rho_{xx}^2$ in the whole temperature range of 2-350 K for all the differently disordered films, well consistent with the scaling $\rho_{AH}=a_0\rho_{xx0}+b\rho_{xx}^2$. In Fig. 3(d), we further plotted anomalous Hall conductivity $\sigma_{AH}$-$\sigma_{xx}^2$ curves measured at different temperature from 2 to 350 K for the $Mn_{1.5}Ga$ films with different $S$, in which the good linearity confirms the good agreement between the measured data and the scaling $\sigma_{AH}=a_0\sigma_{xx0}^{-1}\sigma_{xx}^2+b$. These observations provide strong evidence that the extrinsic contributions to skew scattering from phonons, magnons and weak localization are totally negligible in the case of $Mn_{1.5}Ga$ films, leaving defect-induced skew scattering and intrinsic contribution as the only important origination of $\rho_{AH}$. This experimental observation of negligible skew scattering caused by weak localization are also supported by the early theoretical work.[25,29-30] The discrepancies and puzzlingly large $n$ in Figs. 3(a) and (b) should be attributed to their unreasonable presupposition $\alpha_0=\alpha_L=\alpha_p=\alpha_m$, i.e. the neglect of the qualitative difference between the various possible contributions to extrinsic AHE. We claim that the large $n$ in homogenous films (at least) does not necessarily linked with boundary (surface and interface) scattering.[14] This finding is also inspiring for understanding the interesting scaling behaviors of AHE in other material systems, especially the "puzzling" $n$ below 1 or beyond 2.

The scaling constants $a_0$ and $b$ for differently disordered films are extracted from the best fitting of $\rho_{AH}=a_0\rho_{xx0}+b\rho_{xx}^2$ and $\sigma_{AH}=a_0\sigma_{xx0}^{-1}\sigma_{xx}^2+b$ to the data. As shown in Fig. 4(a), the values of both $a$ and $b$ determined from $\sigma_{AH}$-$\sigma_{xx}^2$ curves coincide with that determined from $\rho_{AH}$-$\rho_{xx}^2$ curves, which confidently supported the scaling totally neglected the skew scattering term related to localization, phonons and magnons. The intrinsic parameter $b$ remarkably increases from $7.83\times10^{-6}$ to $8.65\times10^{-5}$ $\mu\Omega^{-1}$.cm$^{-1}$ as $S$ increases. It has been pointed out that chemical disorder and composition could have pronounced influence on the topology of Fermi surface,[5,18-19] therefore, we refer that the strong dependence of intrinsic term $b$ on $S$ for our $Mn_{1.5}Ga$ films should be attributed to the variation of Fermi surface topology arising from chemical disorder effects. It could be helpful to theoretically repeat the disorder effects on the topology of Fermi surface and intrinsic AHE if a convincing band calculation available. Noticeably, similar to the case of magnetization and coercivity,[23] the extrinsic parameter $a_0$ exhibit a non-monotonic dependence on $S$. It is also not a linear function of $\rho_{xx0}$ for $Mn_{1.5}Ga$ films with different degree of chemical ordering, quite different from the case of Fe films with different thickness,[3] which suggests the strong effects of chemical disorder on both skew scattering ($\alpha_0$) and side jump ($\beta$). Figure 4(b) compares the intrinsic AHE conductivity $\sigma^{int}=b$ and defect-induced extrinsic part $\sigma^{ext}=a_0\sigma_{xx0}^{-1}\sigma_{xx}^2$ for all the films. By comparing the data between films with different $S$, one can find the strong chemical disorder effects on the AHE conductivity. At low $T$, $\sigma^{ext}$ is very important and larger than $\sigma^{int}$ for high $S$ samples; $\sigma^{ext}$ is very small and $\sigma^{int}$ dominates for low $S$ samples. At each fixed finite temperature, as $S$ decreases, $\sigma^{int}$ decreases monotonically, while $\sigma^{ext}$ does not, due to the non-monotonic dependence of $a_0$ and $\rho_{xx}$ on $S$ (or $\rho_{xx0}$). The decrease of $\sigma^{int}$ suggests the suppression of intrinsic contributions to AHE conductivity by chemical disorder. For all the films with different $S$, $\sigma^{ext}$ decreases monotonically as $T$ increases, indicating that the finite-temperature effects greatly suppress the extrinsic contributions and leave the scattering-independent intrinsic contributions as dominate at high $T$, coinciding with previous observations.[2,3]

In conclusion, we have investigated the transport behaviors of weak localization and AHE in perpendicularly magnetized $L1_0$-$Mn_{1.5}Ga$ films. We observed significant weak localization and metal-insulator transition in highly disordered films. We also find that the scaling $\rho_{AH}=a_0\rho_{xx0}+b\rho_{xx}^2$ excellently describes the measured data, while the conventional scaling laws $\rho_{AH}\sim\rho_{xx}^n$ and $\rho_{AH}=a\rho_{xx}+b\rho_{xx}^2$ show significant discrepancies from the experimental data and puzzlingly large $n$ of 2.13. These results provide firm evidence that the extrinsic AHE resistivity induced by weak localization, phonons and magnons is negligible compared to defect-induced part in $Mn_{1.5}Ga$ films. The chemical disorder is found to strongly affect both intrinsic and extrinsic $\sigma_{AH}$. For all the chemical disordered films, the finite-temperature effects greatly suppress the extrinsic AHE conductivity and leave the intrinsic part as dominate at high $T$. Our results help better understand the interesting chemical disorder-induced weak localization, and the chemical disorder effects and finite-temperature effects (weak localization, phonons and magnons) on the scaling behaviors of AHE. Our results offer a technological engineering approach of the AHE by altering long-range chemical ordering parameter of magnetic materials.

We acknowledge G. Y. Guo, Q. Niu for the useful discussion on the results, J. Yan, F. Pang, S. L. Wang, K. K. Meng, J. Lu, X. Z. Yu, H. L. Wang and J. X. Xiao for the help on the sample preparation and transport measurements, X. F. Han, D. L. Li and Q. T. Zhang for the assistance on ion-beam etching during Hall device fabrications. We thank Q. Cai at 4B9A beamline of BSRF, G. Q. Pan at the U7B beamline of NSRL and C. Z. Liu at the BL14B1 beamline of SSRF for their help on the Synchrotron XRD and XRR measurements. This work is supported by NSFC for Grant 61334006 and 11127406, and MOST of China for Grant 2013CB922303.

**Figure captions**

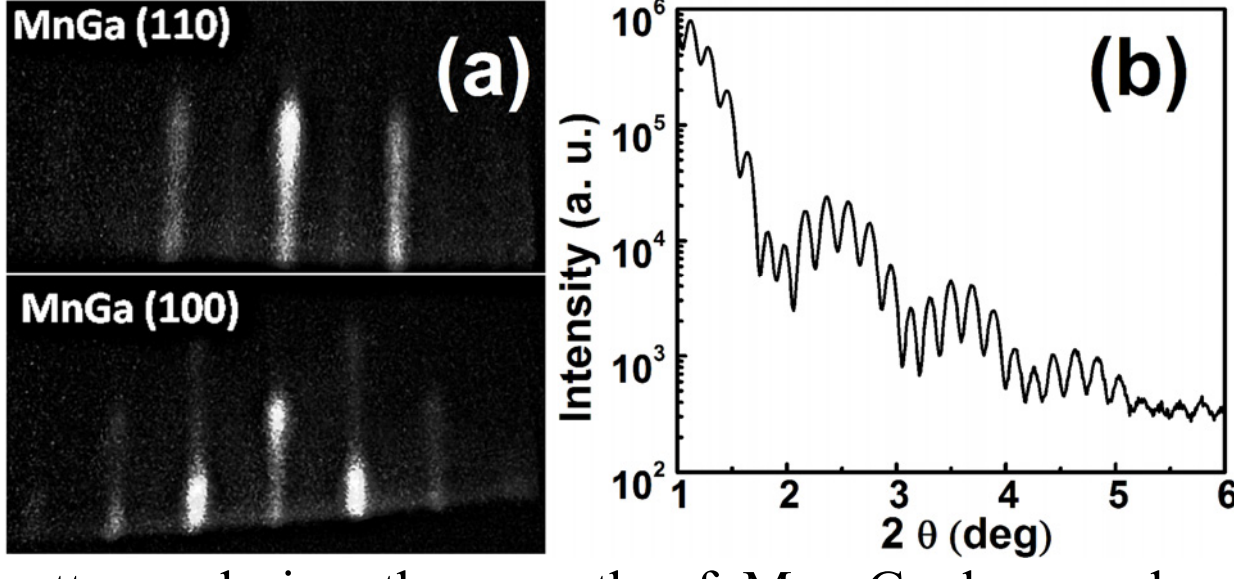

FIG.1 (a) Typical RHEED patterns during the growth of $Mn_{1.5}Ga$ layers along (110) and (100) azimuths, respectively. Along (110) azimuth a weak twofold reconstruction is observed. (b) Representative XRR curve of GaAs (001) /$Mn_{1.5}Ga$ 50 nm/GaAs 6 nm samples.

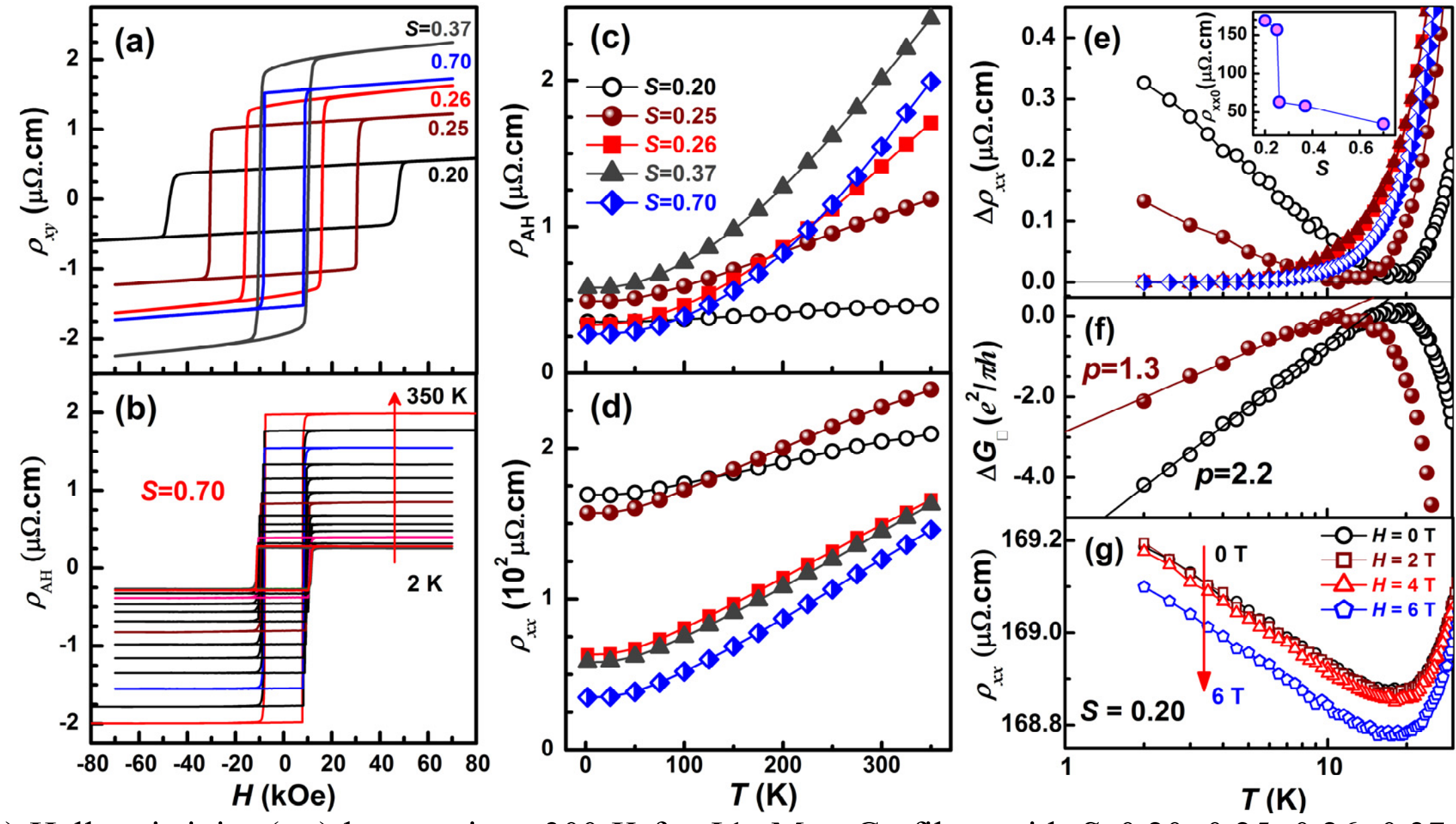

FIG.2 (a) Hall resistivity ($\rho_{xy}$) hysteresis at 300 K for $L1_0$-$Mn_{1.5}$Ga films with *S*=0.20, 0.25, 0.26, 0.37 and 0.70, respectively. Temperature dependence of (b) Anomalous Hall resistivity ($\rho_{AH}$) hysteresis (*S*=0.70), (c) $\rho_{AH}$, (d) $\rho_{xx}$ (*H*=0 T), (e) $\Delta\rho_{xx}$ (*H*=0 T), (f) $\Delta G_\square$ (*S*=0.20 and 0.25, *H*=0 T) and (g) $\rho_{xx}$ (*S*=0.20, ***H***=0, 2, 4 and 6 T) for $L1_0$-$Mn_{1.5}$Ga films. The inset in (*d*) shows the minimum resistivity $\rho_{xx0}$ as a function of *S*. The solid lines in (f) are fits of $\Delta G_\square = pe^2/\pi h \ln(T/T_0)$ with *p* is the slope.

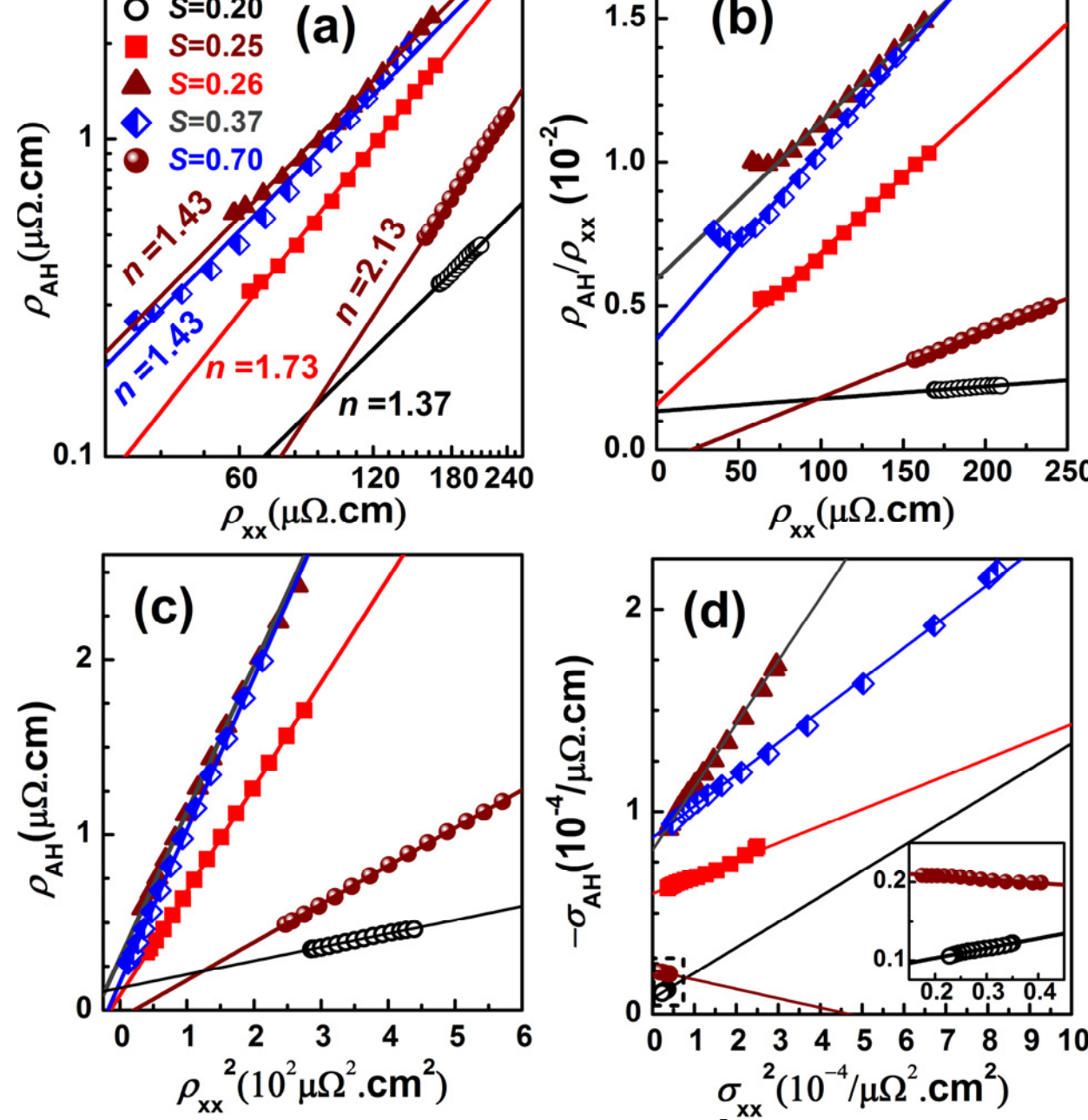

FIG.3 (a) $\rho_{AH}$ versus $\rho_{xx}$, (b) $\rho_{AH}/\rho_{xx}$ versus $\rho_{xx}$, (c) $\rho_{AH}$ versus $\rho_{xx}^2$, and (d) $\sigma_{AH}$ versus $\sigma_{xx}^2$ curves for $L1_0$-$Mn_{1.5}$Ga films with *S*=0.20, 0.25, 0.26, 0.37 and 0.70. The lines are fits with (a) $\rho_{AH}=C\rho_{xx}^{n}$, (b) $\rho_{AH}/\rho_{xx}=a+b\rho_{xx}$, (c) $\rho_{AH}=a_0\rho_{xx0}+b\rho_{xx}^2$, and (d) $\sigma_{AH}=a_0\sigma_{xx0}^{-1}\sigma_{xx}^2+b$, respectively. The inset in (d) is zoom-in plot of the dashed box.

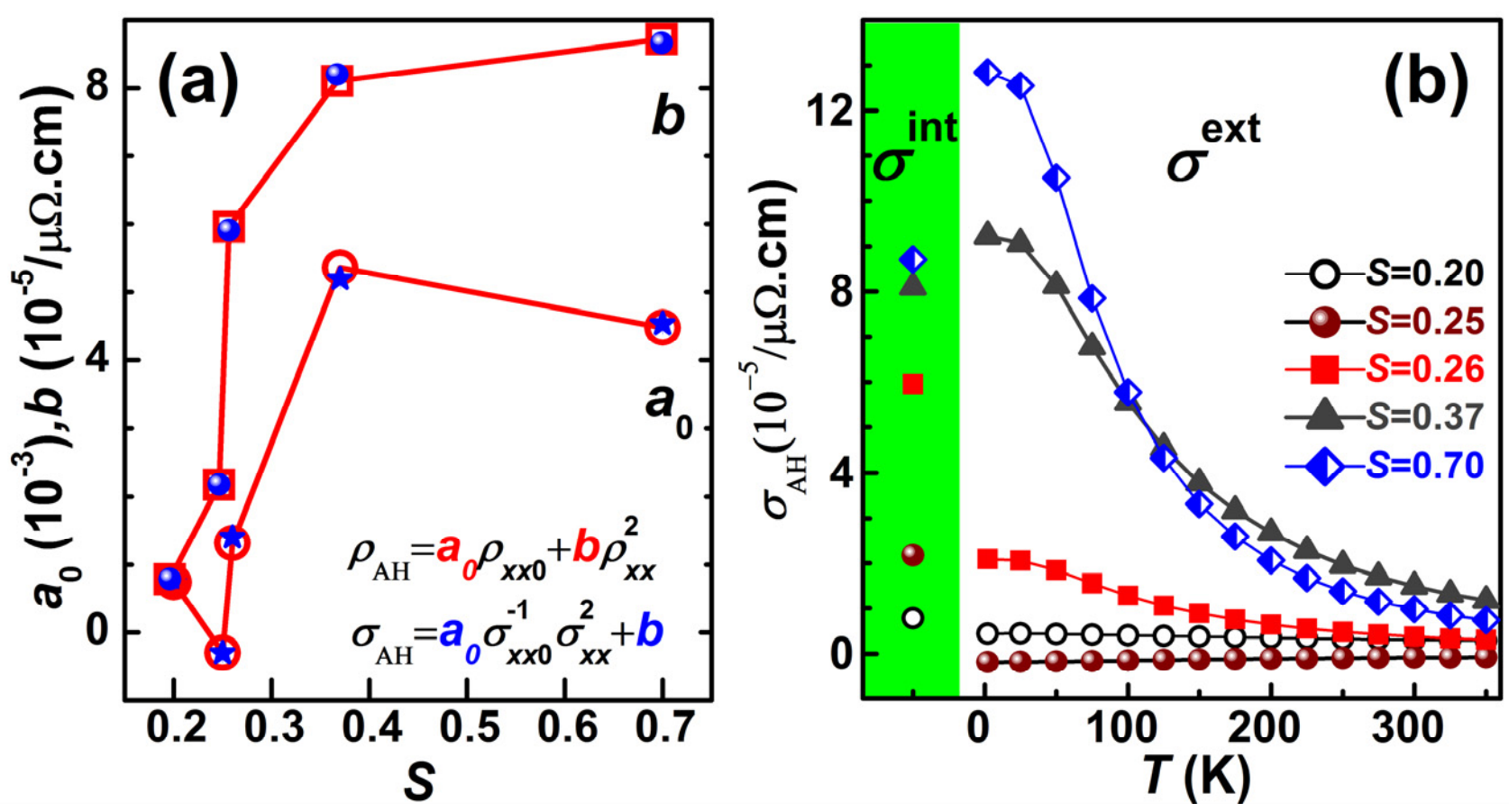

FIG.4 (a) $S$ dependence of $a_0$ and $b$ for $L1_0$-$Mn_{1.5}Ga$ films. The red rectangle (circle) for $b$ ($a_0$) determined from determined by the best fits of the data with $\rho_{AH} = a_0\rho_{xx0} + b\rho_{xx}^2$; blue dot (star) for $b$ ($a_0$) determined from and $\sigma_{AH} = a_0\sigma_{xx0}^{-1}\sigma_{xx}^2 + b$; (b) Temperature-dependent $\sigma^{ext}$ (white regime) and $\sigma^{int}$ (green regime) for $Mn_{1.5}Ga$ films with $S$ = 0.20, 0.25, 0.26, 0.37 and 0.70, respectively.